\documentclass[twocolumn,showpacs,preprintnumbers,amsmath,amssymb]{revtex4}

\usepackage{graphicx}
\usepackage{dcolumn}
\usepackage{bm}
\usepackage{multirow}

\newcommand{\Cl}{$\kappa$-(ET)$_2$Cu[N(CN)$_2$]Cl}
\newcommand{\kClstar}{$\kappa$-Cl$^\ast$}
\newcommand{\kCl}{$\kappa$-Cl}

\begin{document}

\preprint{APS/123-QED}

\title{Energy scales for electronic noise processes in the quasi-two-dimensional organic Mott system $\kappa$-(BEDT-TTF)$_2$Cu[N(CN)$_2$]Cl}
\author{Jens M\"uller}
 \email{j.mueller@physik.uni-frankfurt.de}
\affiliation{Physikalisches Institut, Goethe-Universit\"at Frankfurt am Main, Max-von-Laue-Str.\ 1, 60438 Frankfurt am Main, Germany}
\author{Jens Brandenburg}
\affiliation{Max Planck Institute for Chemical Physics of Solids, N\"othnitzer-Str.\ 40, 01187 Dresden, Germany}
\author{John A.\ Schlueter}
\affiliation{Argonne National Laboratory, Materials Science Division, Argonne, IL 60439, USA}

\date{\today}

\begin{abstract}
Resistance noise spectroscopy is applied to bulk single crystals of the quasi-two-dimensional organic Mott insulator $\kappa$-(BEDT-TTF)$_2$Cu[N(CN)$_2$]Cl both under moderate pressure and at ambient-pressure conditions. When pressurized, the system can be shifted to the inhomogeneous coexistence region of antiferromagnetic insulating and superconducting phases, where percolation effects dominate the electronic fluctuations [J.\ M\"uller {\it et al.}, Phys.\ Rev.\ Lett.\ {\bf 102}, 047004 (2009)]. Independent of the pressure conditions, at higher temperatures we observe generic $1/f^\alpha$-type spectra. The magnitude of the electronic noise is extremely enhanced compared to typical values of homogeneous semiconductors or metals. This indicates that a highly inhomogeneous current distribution may be an intrinsic property of organic charge-transfer salts. The temperature dependence of the nearly $1/f$ spectra can be very well described by a generalized random fluctuation model [Dutta, Dimon, and Horn, Phys.\ Rev.\ Lett.\ {\bf 43}, 646 (1979)]. We find that the number of fluctuators and/or their coupling to the electrical resistance depends on the temperature, which possibly relates to the electronic scattering mechanisms determining the electrical resistance. The phenomenological model explains a pronounced peak structure in the low-frequency noise at  around 100\,K, which is not observed in the resistivity itself, in terms of the thermally-activated conformational degrees of freedom of the ET molecules' ethylene endgroups.
\end{abstract}


\maketitle
\section{Introduction}
In recent years, molecular materials have provided unprecedented model systems for exploring the physics of correlated charge carriers in reduced dimensions \cite{Ishiguro1998,Toyota2007}. For the family of quasi-two-dimensional organic conductors $\kappa$-(BEDT-TTF)$_2$X, where BEDT-TTF (ET in short) is bis(ethylenedithio)tetrathiafulvalene and X stands for polymeric anions, the combined effect of strong electron-electron and electron-phonon interactions, together with the reduced dimensionality, gives rise to a rich phenomenology of ground states like Mott insulating states, which may show long-range antiferromagnetic order or remain a disordered spin liquid down to low temperatures, 
unusual metallic phases, or superconductivity. The materials are layered systems with alternating conducting layers, where ET dimers form an anisotropic triangular lattice, and thin insulating anion layers. The material's actual ground state delicately depends on the interdimer transfer integrals, which determine the bandwidth $W$ and the on-site Coulomb repulsion $U$, and can be fine-tuned either by altering the chemical composition of the anions X or the ET molecules (e.g., by step-wise deuteration, i.e.\ replacing hydrogen by deuterium in the terminal ethylene moieties), or by changing the pressure conditions. Thus, the material with X = Cu[N(CN)$_2$]Br is a superconductor with $T_c = 11.2$\,K, whereas the compound X = Cu[N(CN)$_2$]Cl (\kCl\ in short) is an antiferromagnetic insulator with $T_N = 27$\,K. The latter material in turn shows bulk superconductivity with $T_c = 12.8$\,K upon a moderate pressure of $P \sim 300$\,bar, which demonstrates the universality of the phase diagram of these materials. Pressure studies of \kCl\ revealed a first-order metal-to-insulator transition (MIT) line $T_{MI}(P)$ \cite{Lefebvre2000,Limelette2003,Fournier2003,Kagawa2004,Ito1996}, indicative of a bandwidth-controlled Mott transition \cite{Kanoda1997a,Kino1995}, i.e.\ the opening of a gap in the charge carrying excitations due to electron-electron interactions. For the title compound, this goes along with a ground state characterized by long-range antiferromagnetic ordering of localized moments. In the vicinity of the MIT, a spatially inhomogeneous coexistence of the antiferromagnetic insulating and superconducting phases develops at low temperatures. 

Fluctuation spectroscopy has been used as a powerful method to investigate the intrinsic dynamics of carriers of a large variety of magnetic \cite{Raquet2000},  semiconducting \cite{Mueller2006} and metallic/superconducting materials \cite{Testa1988,Lee1989}, in particular systems close to a MIT \cite{Cohen1994,Kar2003,Jaroszinski2004} or the percolation limit \cite{Garfunkel1985}, see also \cite{Kogan1996} for an overview.
In a recent paper, we reported on studies on the pressure-induced spatially inhomogeneous coexistence state in \Cl\ using resistance noise spectroscopy \cite{Mueller2009}. We quantitatively investigated the nature of percolation at the superconducting transition. In the present study, we aim to use the technique in order to gain information on the relevant energy scales for the electronic scattering processes in the temperature range up to room temperature. Our results, comparing pressurized \Cl\ to the system under ambient conditions, demonstrate that measuring electronic fluctuations is a promising new approach to study the dynamical properties of the correlated charge carriers in molecular materials. A simple phenomenological random fluctuation model can explain the temperature dependence of the nearly $1/f$-noise remarkably well, and allows us to determine the energy distribution of certain fluctuations leading to electronic scattering processes, e.g.\ those related to the 
rotational degrees of freedom of the ET molecules' terminal ethylene groups. The latter undergo a glass-like transition, which has been studied in some detail by thermodynamic methods, see \cite{Toyota2007} for an overview, and is related to  {\it intrinsic} and -- in certain limits -- {\it controllable} disorder, which is known to affect the actual ground-state properties, see e.g.\ \cite{Su1998,Kawamoto1998}. Despite the fact that there is no apparent signature in the samples' resistance curves, we find independent spectroscopic evidence for the excitation of the ethylene endgroups' rotational degrees of freedom from a pronounced peak in the normalized resistance noise.

\section{Experiment}
\begin{figure}[]
\includegraphics[width=.475\textwidth]{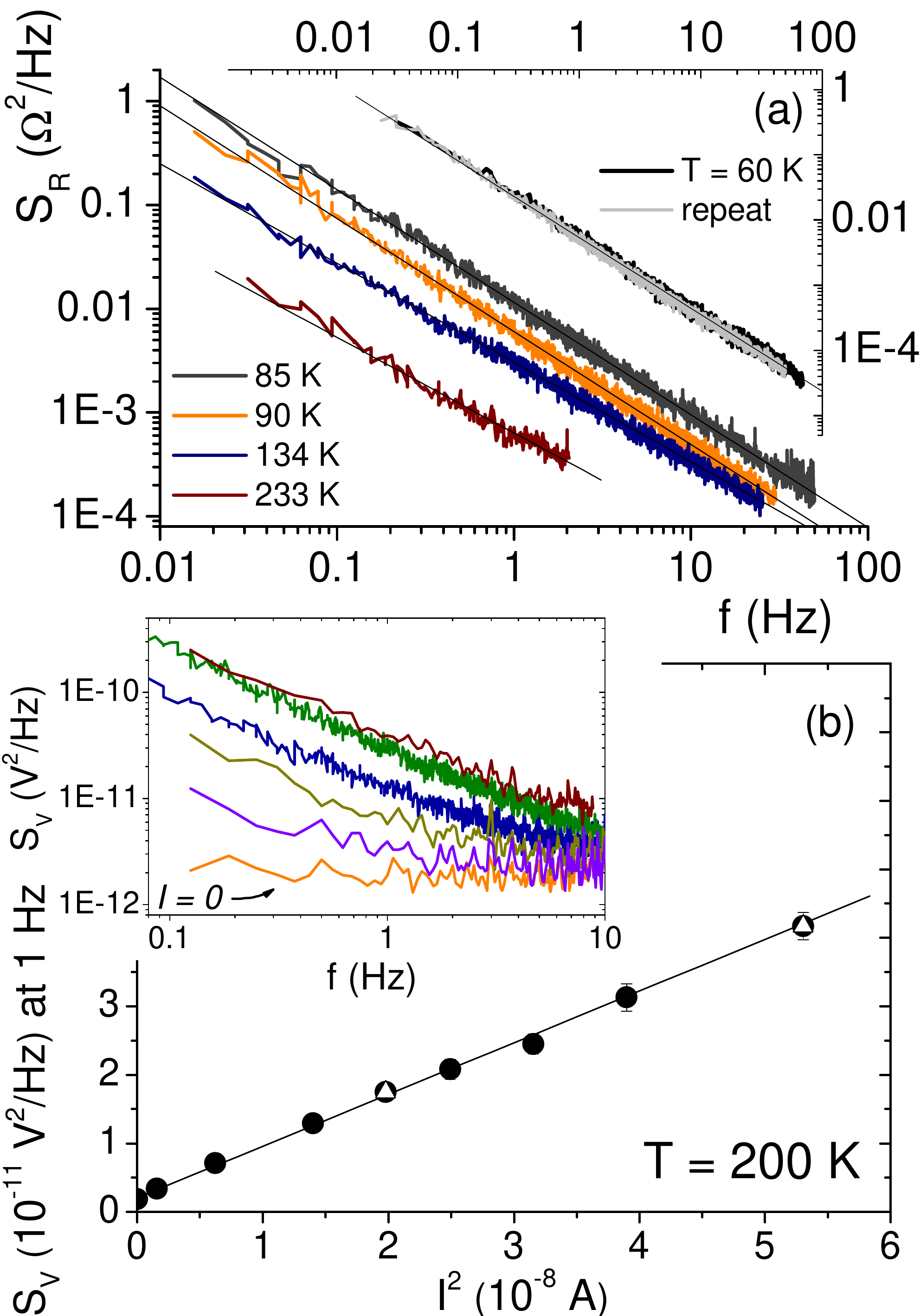}
\caption[]{(Color online). (a) Resistance noise power spectral density $S_R (f)$ of a pressurized sample of $\kappa$-(ET)$_2$Cu[N(CN)$_2$]Cl (denoted as \kClstar) at different temperatures in a log-log plot. Lines are linear fits yielding a power-law behavior $S_R (f) \propto 1/f^\alpha$. The two spectra taken at 60\,K demonstrate an excellent reproducibility. (b) Voltage noise power spectral density $S_V (f)$ taken at $f = 1$\,Hz at $T = 200$\,K for different bias currents $I$. The line is a linear fit to the data in the representation $S_V$ vs $I^2$. Open triangles represent spectra obtained for a different value of the current limiting resistor $r$ yielding an identical scaling behavior. Inset shows the corresponding spectra $S_V (f)$ for different bias currents $I$. Note a "white" floor noise spectrum for $I = 0$.}\label{figure1}
\end{figure}Single crystals of \Cl\ were grown by electrochemical crystallization as described elsewhere \cite{Williams1990}. Samples 
were of plate-like (smallest dimension perpendicular to the conducting planes) and rod-like (longest dimension perpendicular to the planes) morphology, with typical dimensions of $(0.6 \times 0.8 \times 0.2)\,{\rm mm}^3$ and $(0.36 \times 0.36 \times 1.2)\,{\rm mm}^3$, respectively. Electrical contacts for transport measurements have been made using Carbon paste or by evaporating Gold and subsequent contacting with Silver paint. The latter procedure allowed for well-defined contact geometries by using shadow masks. The electrical contacts showed linear $I$-$V$-characteristics and contact resistances in the order of $1 - 10\,\Omega$. As described in \cite{Mueller2009}, we produce a small pressure by embedding samples (denoted as $\kappa$-Cl$^\ast$ hereafter) in a solvent-free epoxy, the slightly larger coefficient of thermal expansion of which results in a finite stress acting on the sample during cooldown. An estimated effective pressure of $220 - 250$\,bar on our samples agrees well with a comparison to resistivity curves obtained from He-gas pressure experiments reported in the literature \cite{Kagawa2004} and the effect on a superconducting reference compound for which the pressure dependence of $T_c$ is known. In the present paper, we discuss a pressurized sample $\kappa$-Cl$^\ast$ exhibiting an inhomogeneous coexistence of insulating and superconducting phases at low temperatures (same sample as in \cite{Mueller2009}) in comparison to the compound under ambient-pressure conditions, \kCl, having an antiferromagnetic insulating ground state. For the latter, due to the size and shape of the samples available, only four-terminal measurements using Carbon-paste contacts were possible. We found a strong increase of the contact resistances and non-ohmic $I$-$V$-characteristics below about 20\,K. Therefore, we will only discuss measurements on \kCl\ performed at temperatures higher than that.\\ 
Low-frequency resistance fluctuations have been measured by a standard bridge-circuit ac-technique \cite{Scofield1987} in a 
five-terminal (\kClstar) or four-terminal (\kCl) setup. The output signal of a lock-in amplifier (SR830), operating at a driving frequency $f_0$ of typically $237 - 517$\,Hz, was processed into a spectrum analyzer (HP35660A) by means of a low-noise preamplifier (SR554 or SR560), which is preferably operated near the eye of its noise figure. Care was taken to rule out or eliminate spurious noise sources, especially those coming from the electrical contacts. The observed noise spectra are found to be independent of the lock-in amplifier, its driving frequency, and the preamplifier as long as the conditions for proper impedance matching are met. In our ac-circuit, the 90$^\circ$ phase-shifted ({\it y}-channel) signal showed frequency-independent ("white") noise and -- representing the noise floor of the experimental setup -- could be subtracted from the {\it x}-channel signal. Figure\,\ref{figure1} (a) shows typical resistance noise power spectral densities (PSDs) of \kClstar\ taken at different temperatures. At all temperatures discussed here, we observe an excess noise spectrum of generic $1/f^\alpha$-type, for which we find $0.9 < \alpha < 1.1$ (see also Fig.\,\ref{figure3}). Lines are fits to the data yielding the magnitude of the noise (spectral weight) and the frequency exponent $\alpha$. Also shown are two spectra taken at the same temperature of 60\,K, where the sample's temperature had been changed in between the measurements. The excellent reproducibility of the spectra allows us to perform a detailed analysis of the temperature dependence of the resistance fluctuations as discussed below. Figure\,\ref{figure1} (b) shows the measured voltage noise PSDs, $S_V (f )$, taken at a temperature of 200\,K for different bias currents $I$ (inset) and the values taken at 1\,Hz, $S_V (f = 1\,{\rm Hz})$. We observe an excellent scaling $S_V \propto I^2$ as expected when the measured noise is solely due to {\em resistance} fluctuations. Also, as expected and required, the observed scaling is independent of the ratio $R/r$, where $R$ is the sample resistance and $r$ the value of the current limiting resistor in our bridge circuit. Furthermore, as shown in the inset, a "white" floor noise spectrum ({\it x}-channel signal) is observed for $I \rightarrow 0$.

\section{Results and Discussion} 
\begin{figure}[]
\includegraphics[width=.475\textwidth]{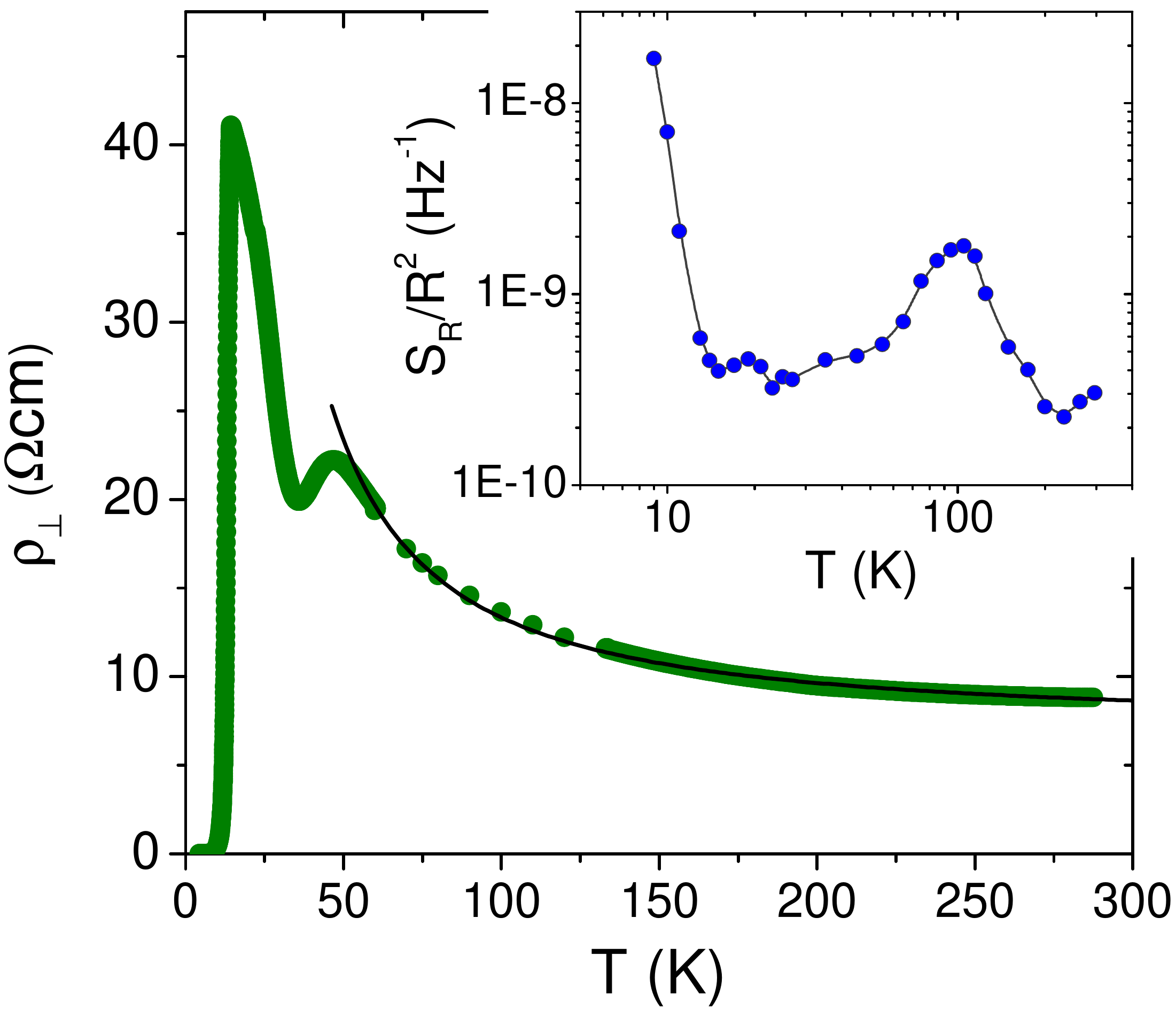}
\caption[]{(Color online.) Resistivity of \kClstar\ located in the vicinity of the first-order Mott metal-to-insulator transition (see Fig.\,1 in \cite{Mueller2009}). The solid line is a fit to the data at temperatures above the transition into the metallic state to $\rho (T) = \rho_A \exp(- \Delta/T) + \rho_B T^{-3/2} $, see text. Inset shows the temperature dependence of the normalized noise $S_R/R^2(T)$ taken at 1\,Hz.}\label{figure2}
\end{figure}
Figure\,\ref{figure2} shows the temperature dependence of the resistivity of a pressurized sample \kClstar. The sample's position in the phase diagram is shifted towards the metallic side such that the S-shaped transition line of the MIT is crossed twice when lowering the temperature (see Fig.\,1 in reference \cite{Mueller2009}). At elevated temperatures, the sample shows a semiconducting behavior ($d\rho/dT > 0$) down to about 46\,K 
below which the sample shows a pronounced minimum following the transition into the metallic phase ($d\rho/dT > 0$). Upon further cooling, the first-order MIT line is crossed again leading once more to insulating behavior below about 22\,K, where a kink in the resistivity curve is observed. Finally, the sample undergoes an insulator-to-superconductor transition at $T_c = 13$\,K. The behavior is qualitatively in good agreement with the phase diagram determined from isobaric temperature and isothermal pressure sweeps by Kagawa {\it at al.} \cite{Kagawa2004,Kagawa2005}. The fact that the transitions are broadened and the transition temperatures are somewhat higher compared to literature results \cite{Ito1996,Lefebvre2000,Kagawa2004} is most likely due to slightly non-hydrostatic pressure conditions in the present experiment. The inset of Fig.\,\ref{figure1} shows the temperature dependence of the normalized noise power $S_R/R^2$ taken at 1\,Hz for \kClstar\ as derived from spectra like the ones shown in Fig.\,\ref{figure1}. The behavior of this sample will be compared to results at ambient pressure in the final part of this section. The reference samples \kCl\ measured under these conditions showed the expected semiconducting behavior down to low temperatures, see Fig.\,\ref{figure6}.

\subsection{Magnitude of the noise}
Before we come to a detailed analysis of the temperature dependence of the noise, we comment on the magnitude of the excess electronic fluctuations. For local and independent noise sources, the spectral density ordinarily scales inversely with system size, which may be expressed by the empirical Hooge relation $S_R(f)/R^2 = \gamma_H / n_c \Omega f$ \cite{Hooge1969,Weissman1988}, where $\gamma_H$ denotes the Hooge parameter representing the strength of the normalized $1/f$-noise, $n_c$ the charge carrier density, and $\Omega$ the noisy volume. In order to get a rough estimate for $\gamma_H$ in the present charge-transfer salts $\kappa$-(ET)$_2$X, we take a carrier density of $n_c \sim 1.2 \times 10^{21}$\,cm$^{-3}$, which results from the present stoichiometry (two ET molecules transfer one electron to the monovalent anion) and for $\Omega$ the total sample volume of $\sim 1.6 \times 10^{-4}$\,cm$^{3}$. Using these parameters we find a room-temperatre value of $\gamma_H$ in the order of $10^{7}$. This number is several orders of magnitude larger than values of $\gamma_H \sim 10^{-2} - 10^{-3}$ typically found for homogeneous metals and semiconductors \cite{Weissman1988}. It is worth noting that such extremely high noise levels also have been found for a number of high-$T_c$ cuprate superconductors \cite{Testa1988,Lee1989,Maeda1989,Maeda1991,Song1991}. For these materials, however, it has been pointed out \cite{Kogan1996} that the high noise level is not inherent to the class of substances but can be explained, e.g.\ by the granularity of the samples, and can thus be found to be largely reduced in samples (in particular thin films) with better crystalline quality and less disorder. For the present molecular materials, which are in the clean limit \cite{Toyota2007}, an explanation in terms of the crystalline quality is not necessarily obvious. Instead, a number of other explanations might account for the high noise level. 
For instance, the noise parameter $\gamma_H$ in the Hooge formula might be overestimated by an incorrect estimate for the carrier density $n_c$. This possibility, however, seems unlikely since for the related metallic system $\kappa$-(ET)$_2$Cu(NCS)$_2$, for which our assumption for $n_c$ can be considered as realistic, we also find a strongly enhanced value of $\gamma_H \sim 10^{5}$ \cite{comment1}. Alternatively, the noise level might be overestimated by an incorrect estimate of the volume $\Omega$ between the electrical probes, e.g.\ due to inhomogeneous current paths in the sample. Since the Hooge expression is empirical, deviations may not be surprising and may point to differences in the conduction mechanism in the present molecular metals and the materials in which it works well \cite{Galchenkov1998}. This explanation includes the possibility that the large noise level could be an intrinsic property of the (ET)$_2$X compounds originating, e.g., in an inhomogeneous current distribution. Such an interpretation indeed has been suggested for explaining the large noise magnitude ($7 - 8$ orders of magnitude higher than expected from the Hooge formula) in bulk crystals of the quasi-one-dimensional organic charge-transfer salt TTF-TCNQ \cite{Galchenkov1998}. Assuming that crystal imperfections force the charge carriers to hop from stack to stack in the process of propagation along the stacking $b$-axis, the authors estimate a number of charge propagation paths, which, in their model, explains the observed enhancement of the noise magnitude. Systematic studies of the noise level in a number of different quasi-two-dimensional organic charge-transfer salts (ET)$_2$X are in progress in order to get a more detailed understanding of this effect.

\subsection{Temperature dependence of the noise}
\subsubsection*{Random fluctuation model}
\begin{figure}[]
\includegraphics[width=.45\textwidth]{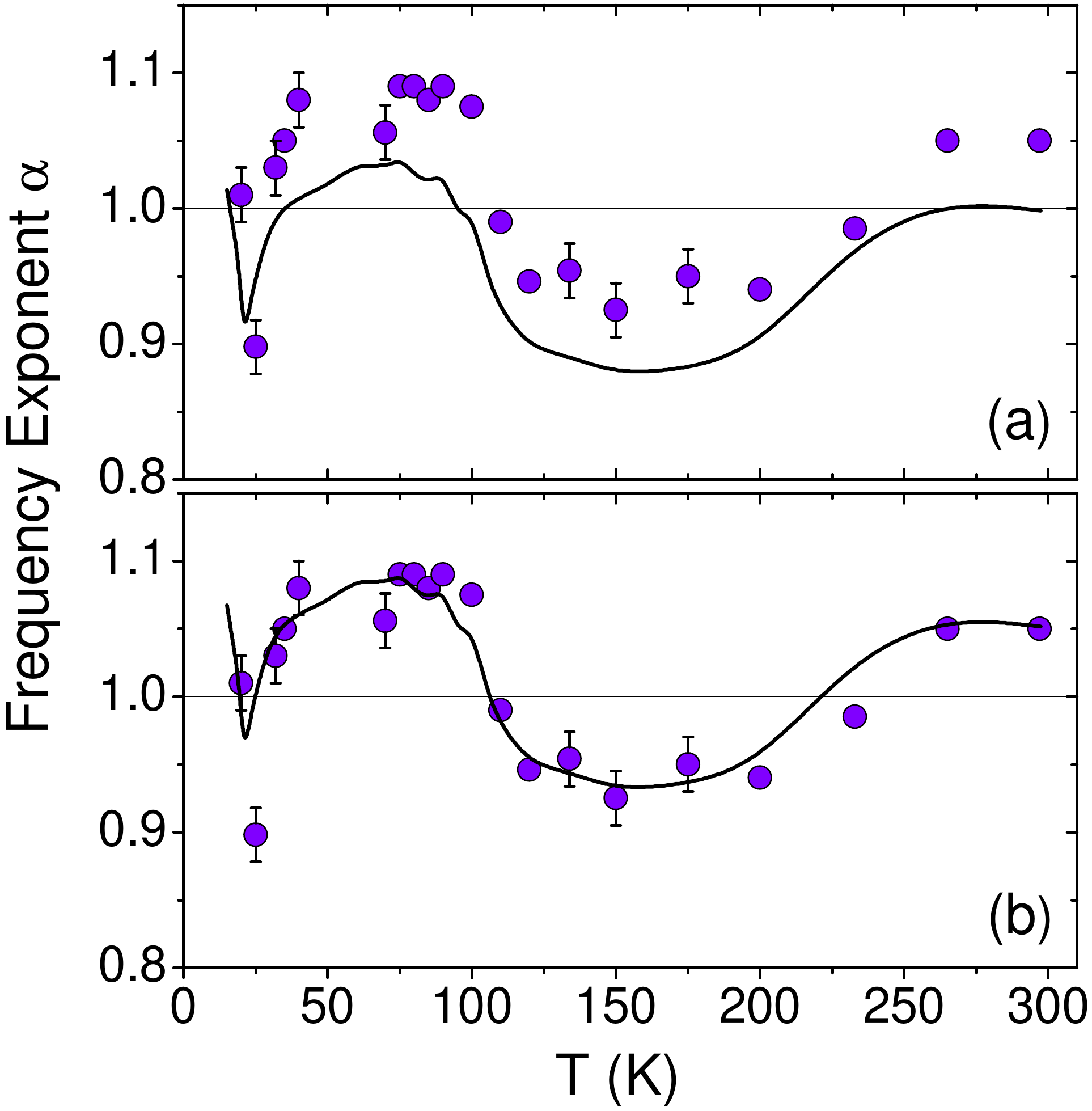}
\caption[]{(Color online). (a) Frequency exponent of the $1/f^\alpha$-noise as a function of temperature, $\alpha(T)$, of \kClstar. Circles are the values extracted from the fits to the individual spectra at different temperatures, see Fig.\,\ref{figure1}. The solid line is the predicted $\alpha(T)$ from the data shown in Fig.\,\ref{figure2} using Eq.\,\ref{DDHeq3} and taking $g(T) = const$. (b) Best fit between the experimental $\alpha(T)$ and the predicted values using Eq.\,\ref{DDHeq3} while allowing for a temperature dependent function $g(T)$ yielding $g(T) = a T^{-3/2}$.}\label{figure3}
\end{figure}

We now turn to the analysis of the temperature dependence of the noise. The inset of Fig.\,\ref{figure2} shows the normalized noise power $S_R/R^2(T)$ taken at 1\,Hz as deduced from the resistance noise PSDs taken at different temperatures, see Fig.\,\ref{figure1}. At present there is no theory for the origin of $1/f$-noise in the low-dimensional molecular metals discussed here. We therefore analyze our data using the phenomenological random fluctuation model introduced by Dutta, Dimon, and Horn (DDH) \cite{Dutta1979}, who showed that noise data in inhomogeneous metals can be transformed to obtain the distribution of activation energies for the processes that generate the noise. The model assumes that the approximate $1/f$-spectra are created by a broad distribution of activation energies of a large number of random and independent switching entities called "fluctuators", which are characterized by thermally-activated time constants $\tau = \tau_0 \exp(E/k_B T)$, where $\tau_0$ is a characteristic Òattempt timeÒ of the order of an inverse phonon frequency, $E$ the activation energy of the process and $k_B$ Boltzmann's constant. The resistance is linearly coupled to the fluctuating quantity, which in this model is not specified {\em a priori} but may be identified {\em a posteriori}, e.g.\ if its energy signature is known. The total noise spectrum can be deduced by integrating over the Lorentzian spectra of the individual fluctuators weighted by a temperature dependent function $g(T)$ and the distribution of activation energies $D(E)$: 
\begin{equation}
S(f,T) \propto \int g(T) \frac{\tau(E)}{\tau(E)^2 4 \pi^2 f^2 +1}D(E)dE.
\label{DDHeq1}
\end{equation}
If $D(E)$ varies slowly compared to $k_B T$, the energy distribution of the fluctuators creating the $1/f^\alpha$-noise may be deduced directly from the measured temperature dependence of the normalized noise by
\begin{equation}
D(E) \propto \frac{2 \pi f S(f,T)}{k_B T} \cdot \frac{1}{g(T)},
\label{DDHeq2}
\end{equation}
where the transition energies are roughly linked to the frequency bandwidth and the temperature \cite{comment2} by 
\begin{equation}
E \approx - k_B T \ln(2 \pi f \tau_0). 
\label{DDHeq4}
\end{equation}
In the original paper, DDH assumed that the net mean-square fractional resistance fluctuations integrated over all frequencies are temperature independent, so in the original model $g(T) = const.$ In this case, the limit $D(E) = const.$ generates exact $1/f$-noise, i.e.\ $\alpha(T) = 1$ and $S_R \propto k_B T / f$.
We will show that in the present case a generalized DDH model \cite{Raquet1999} allowing for a temperature-dependent function $g(T)$, which quantifies the temperature dependence of the noise level, is useful and may give additional insight into the understanding of the relevant scattering processes. An essential ingredient of the DDH model is that a given $D(E)$ is the origin of both the small deviations of $\alpha(T)$ from 1 and the possibly strong dependence of $S_R$ on $T$ \cite{Dutta1979}. Therefore, the frequency exponent of the spectral density and the temperature dependence of the noise are related by:
\begin{equation}
\alpha(T) = 1 - \frac{1}{\ln(2 \pi f \tau_0)} \left[ \frac{\partial \ln S(f,T)}{\partial \ln T} - \frac{\partial \ln g(T)}{\partial \ln T} - 1\right].
\label{DDHeq3}
\end{equation}
Thus, although the DDH model is phenomenological and its assumptions are quite general, the consistency of the latter can be checked independently by comparing the observed $\alpha(T)$ with the ones predicted by Eq.\,(\ref{DDHeq3}) using the temperature dependence of the noise. Figure\,\ref{figure3} shows
the frequency exponent $\alpha(T)$ obtained from the fits to the spectra at different temperatures. We find that the data are closely distributed around $\alpha = 1$ but show a non-monotonic behavior, which reflects the shape of the distribition of the activation energies, i.e.\ for $\alpha$ greater or smaller than 1, $\partial D(E)/\partial E >0$ and $\partial D(E)/\partial E >0$, respectively. The predicted values (solid lines in Fig.\,\ref{figure3}) are in very good agreement with the experimental results: the overall course is well reproduced, which means that the assumptions of the model are consistent and the distribution of energies can be reliably extracted from Eq.\,\ref{DDHeq2}. A striking feature, however, is that there is a constant offset between the experimental and theoretical curve, when we use the original DDH model, in which $g(T) = const.$, i.e.\ when the second term in the bracket of Eq.\,\ref{DDHeq3} vanishes. The clear vertical shift we observe can be taken as evidence that one cannot assume that there is no net temperature dependence of the mean-square fractional resistance fluctuations \cite{Raquet1999}. Since according to Fig.\,\ref{figure3}, $\partial \ln g(T) / \partial \ln T$ in Eq.\,\ref{DDHeq3} is roughly temperature independent, $g(T)$ satisfies a simple power law $g(T) = a T^b$, where the exponent $b$ is unambiguously given by the vertical shift. The best fit to the data yields a parameter $b = -1.5 \pm 0.1$. Following a similar analysis for the $1/f^\alpha$-noise in a magnetic manganite film \cite{Raquet1999}, we conclude that the total noise level is temperature dependent and the number and/or strength of the fluctuators (i.e.\ their coupling to the electrical resistivity) increases with decreasing temperature according to a $T^{-3/2}$ law. It is now interesting to note that this power law can be used as an ingredient to roughly describe the temperature dependence of the sample's resistivity itself. As shown in Fig.\,\ref{figure2}, we have fitted the resistivity data at temperatures above the transition into the metallic state to $\rho (T) = \rho_A \exp(- \Delta/T) + \rho_B T^{-3/2} $ and find a good agreement. The value for the effective charge gap extracted from the fit is $\Delta \sim 30$\,K, which in turn is in good agreement with the values of $\Delta(P)$ found close to the critical pressure of $P_c = 232$\,mbar determined by Kagawa {\it et al.} in resistance measurements at different pressures, see Fig.\,5 of \cite{Kagawa2004}. It is tempting to conclude that the mechanism responsible for the $T^{-3/2}$ term in the electrical resistivity can be linked to the $g(T)$ function via the number and strength of the fluctuators. The same scattering mechanism seems to be responsible for the resistivity and its fluctuations. Theoretical work is highly desirable in order to identify the relevant fluctuations and the coupling mechanism to the electrical resistivity. 
 
\subsubsection*{Distribution of activation energies}
\begin{figure}[]
\includegraphics[width=.475\textwidth]{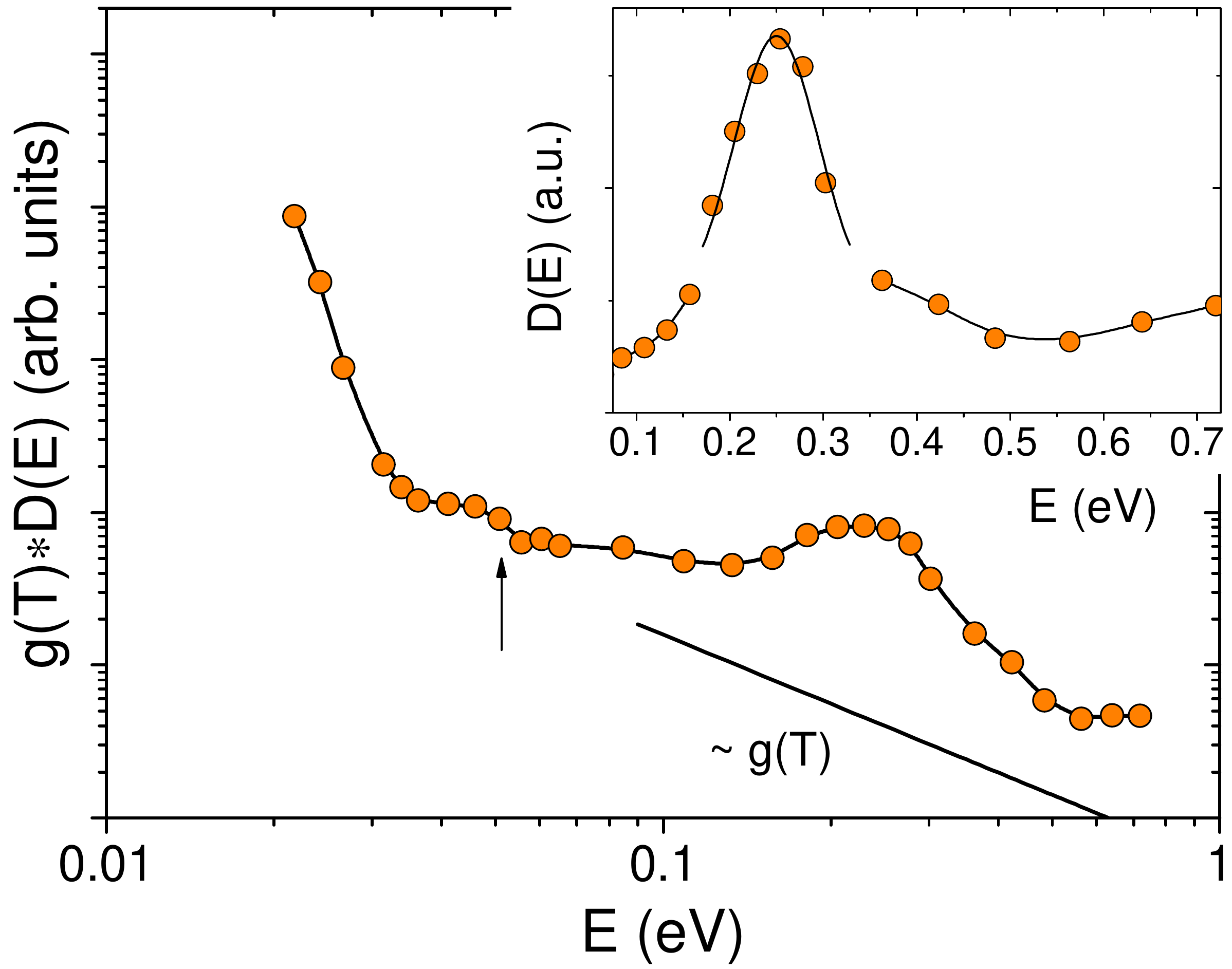}
\caption[]{(Color online). Distribution of activation energies in a log-log plot as $\{ g(T) \ast D(E) \}$ vs $E$ calculated from the data in Fig.\,\ref{figure2}. The function $g(T) \propto T^{-3/2}$ is also shown. The arrow indicates the temperature of reentering the insulating state. Note a pronounced peak in $D(E)$ at $\sim 230$\,meV. Inset depicts $D(E)$ vs $E$ in a linear representation in the vicinity of the peak. The solid line centered about the maximum is a Gaussian fit. }\label{figure4}
\end{figure}

\begin{figure}[]
\includegraphics[width=.475\textwidth]{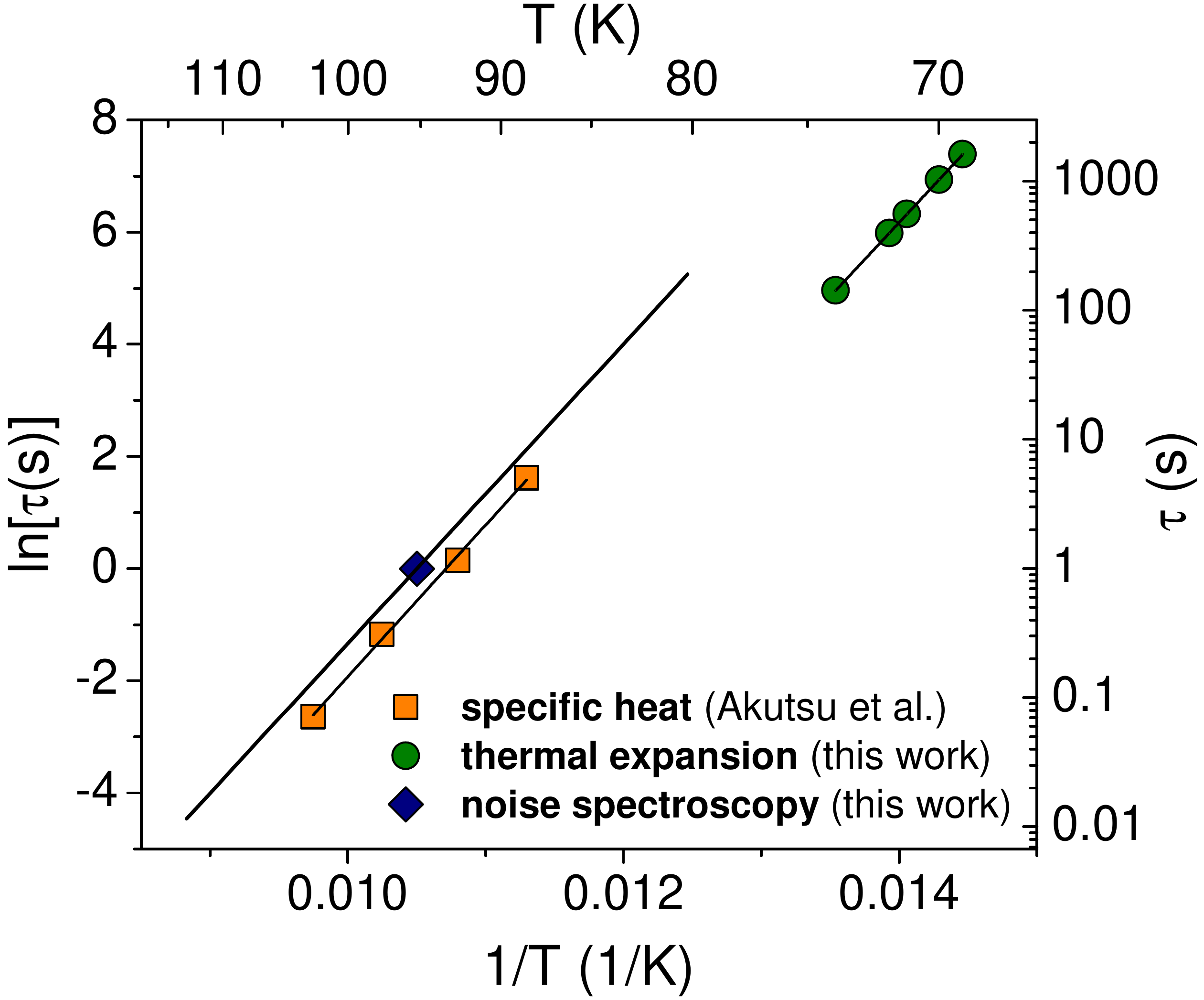}
\caption[]{(Color online). Arrhenius plot of the anomalies in \Cl\  related to the glass-like transition in specific heat (squares, Ref.\,\cite{Akutsu2000}) and thermal expansion (circles). Lines represent activation energies of $E_a = 223$\,meV and 227\,meV, respectively. Also indicated is the fact that resistance fluctuation spectroscopy at 1\,Hz reveals a pronounced peak in $D(E)$ at about 230\,meV, see Fig.\,\ref{figure4}. Length of the line represents the accessible frequency range in the present experiment.}\label{figure5}
\end{figure}

Fig.\,\ref{figure4} shows the energy distribution $D(E)$ extracted from the noise data using the DDH model.
In the main panel, the energy distribution $D(E)$ convolved with $g(T)$ is shown as calculated from the normalized noise data $S_R/R^2(T)$ and using Eqs.\,\ref{DDHeq2} and \ref{DDHeq4} with $\tau_0 = 10^{-13}$\,s. Due to the only logarithmic dependence of the energy on frequency bandwidth  and attempt time, the results are extremely insensitive to the latter \cite{Dutta1979}. The distribution shown in the main panel is identical to the one extracted by the original DDH model, i.e.\ {\it without} considering a temperature-dependent function $g(T)$. One can clearly see that there is a monotonous increase of the density of fluctuators having a characteristic energy $E$ with decreasing energy. As discussed above, a convolution with a function $g(T) \propto T^{-3/2}$ best describes the data. Besides this general trend, we observe three striking features in $D(E)$. (i) The more than two-orders-of-magnitude increase at low energies has been discussed in detail in Ref.\,\cite{Mueller2009} in terms of percolative superconductivity in the inhomogeneous coexistence region of antiferromagnetic insulating and superconducting phases. We have shown that the observed behavior can be understood in the framework of a random-resistor network. (ii) We attribute the small but distinct step-like increase at about 50\,meV to the crossing of the metal-to-insulator transition line when reentering the insulating state. Qualitatively, one expects an increase of the noise  due to localization of the carriers. This interesting aspect of the Mott MIT may be best studied by experiments where the actual pressure can be tuned continuously. Such experiments are in preparation. We now focus on (iii) the pronounced, roughly Gauss-shaped peak at an energy of $230 - 250$\,meV, which corresponds to a pronounced maximum in the normalized noise level at around 100\,K. Interestingly, the energy where $D(E)$ peaks, corresponds to the energy signature of the orientational degrees of freedom of the ET molecules' terminal ethylene moieties undergoing a glass-like transition. The ethylene endgroups can adopt two possible configurations, with the relative orientation of the outer C--C bonds being either eclipsed or staggered. While at high temperatures the ethylene-endgroup system is thermally disordered, such that the two configurations are roughly equally populated, a preferential orientation in the eclipsed configuration is adopted upon lowering the temperature. A kinetic, glass-like transition occurring at a characteristic temperature $T_g$, however, prevents a fully-ordered low-temperature state, and a certain degree of disorder depending on the thermal history becomes frozen. Evidence for a glassy freezing of the ethylene endgroups came from step-like anomalies in the heat capacity around $90 - 105$\,K, whereas the temperature of the anomalies was found to depend on the measurement frequency \cite{Akutsu2000}. From this frequency dependence an activation energy of $E_a = 233$\,meV has been inferred. The authors pointed out that the glass transition plays no important role in the electrical properties for the $\kappa$-(ET)$_2$Cu[N(CN)$_2$]Cl salt. Indeed, there is no anomaly visible in the resistivity at that temperature; the noise properties, however, are extremely sensitive to this effect, see Fig.\,\ref{figure2}. In another thermodynamic study, performed by the present author, the glass-like nature of the transition was demonstrated by observing characteristic anomalies in the linear coefficients of thermal expansion at around $T_g = 75$\,K in $\kappa$-(ET)$_2$Cu[N(CN)$_2$]Br and 70\,K in $\kappa$-(ET)$_2$Cu[N(CN)$_2$]Cl \cite{Mueller2002a,Mueller2004} for a cooling rate of $q \sim 5$\,K/h. From the cooling-rate dependence of the glass-transition temperature $T_g$, we then inferred $E_a = 227 \pm 26$\,meV for the latter salt. 
Figure\,\ref{figure5} shows the temperatures of the glass-like transition $T_g$ as determined from specific heat \cite{Akutsu2000} and thermal expansion measurements in an Arrhenius plot, demonstrating the thermally-activated behavior of the characteristic times $\tau = \tau_0 \exp(E_a/k_B T)$. 
Note that the relaxation times $\tau$ have been determined in a different way for the two experiments \cite{comment3}, which explains the offset, apparently corresponding to different attempt times $\tau_0$. The slopes giving the activation energy $E_a$, however, are in very good agreement. In $^1$H-NMR measurements, a similar value of $224 \pm 9$\,meV 
for the activation energy of the ethylene endgroup vibration has been found \cite{Miyagawa1995}. In this respect, it is evident that the peak at about 230\,meV observed in the energy distribution $D(E)$ of fluctuators causing the nearly $1/f$-noise spectra originates in the conformational motion of the ET molecules' terminal ethylene groups. The coupling of these rotational degrees of freedom to the electronic properties can be directly seen in the electronic fluctuations. The line in Fig.\,\ref{figure5}, the slope of which corresponds to this activation energy, shows the frequency range covered by the present experiment. If we identify the fluctuators causing the enhanced electronic noise at 1\,Hz around 100\,K with the ET molecules' terminal ethylene groups and the peak energy of about 230\,meV as the activation energy for transitions between the eclipsed and staggered conformation, then the observed width of $D(E)$ of $\sim 90$\,meV gives an estimate for the range of activation energies involved. The consequences of such a finite distribution of energy barriers relevant for the glassy behavior are to be investigated in more detail. Resistance noise turns out to be a sensitive probe to further study not only the glassy dynamics itself but possibly the influence on the electronic ground-state properties as well, see also \cite{Mueller2009}.\\ Finally, we note that the temperature of the noise maximum of about 100\,K may suggest that the resistivity hump of yet unknown origin observed in the same temperature range in many metallic (ET)$_2$X compounds (see \cite{Ishiguro1998,Toyota2007} and references therein) could be related to the rotational degrees of freedom of the ET molecules.

\subsubsection*{Results at ambient pressure}
\begin{figure}[]
\includegraphics[width=.475\textwidth]{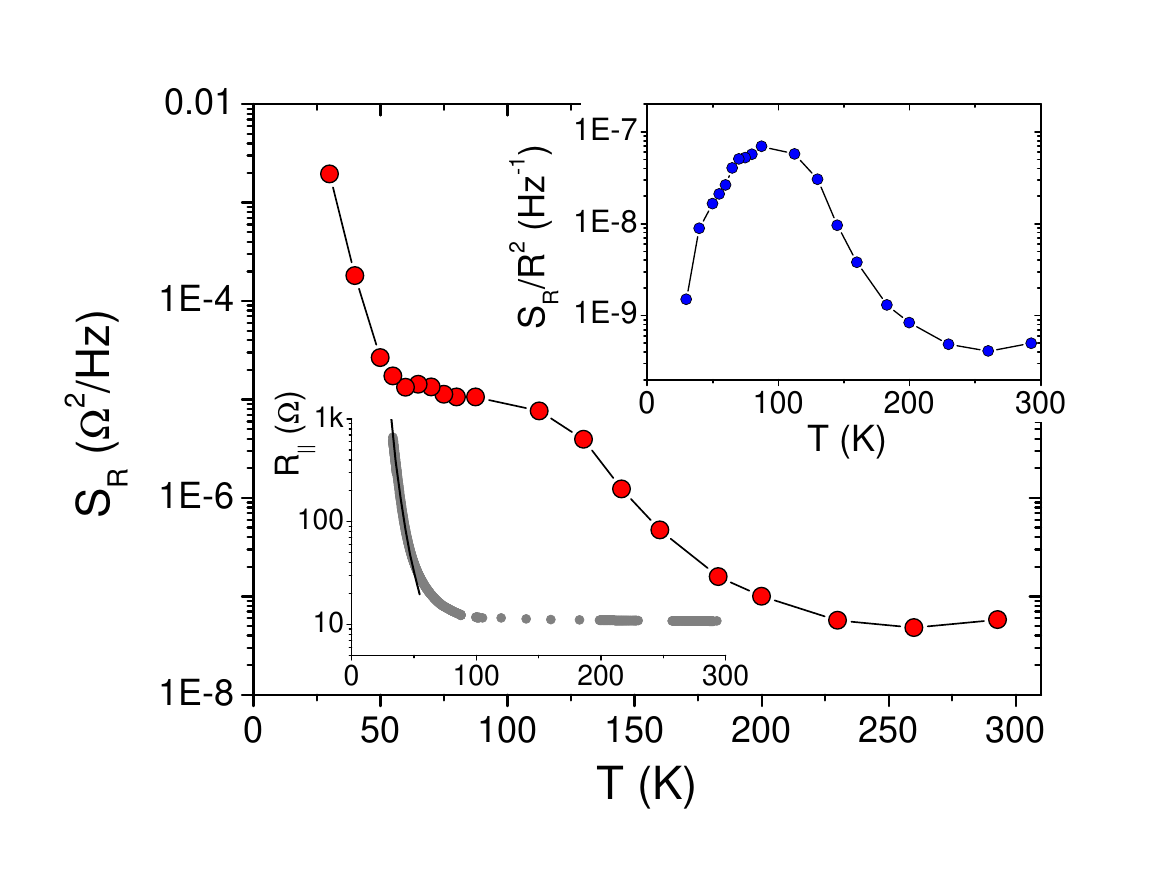}
\caption[]{(Color online). Noise power spectral density $S_R$ (main panel) and normalized noise $S_R/R^2$ (right inset) taken at 1\,Hz of \kCl\ as a function of temperature. Left inset: in-plane resistance showing semiconducting (insulating) behavior. Line is a fit $R \propto \exp(\Delta/T)$ yielding an effective charge gap of $\Delta \sim 300$\,K.}\label{figure6}
\end{figure}
In the last part of this section, we present the data on a reference sample of $\kappa$-(ET)$_2$Cu[N(CN)$_2$]Cl under ambient conditions. Figure\,\ref{figure6} shows the in-plane resistance and noise data as a function of temperature. In contrast to pressurized \kClstar\ 
\begin{figure}[]
\includegraphics[width=.45\textwidth]{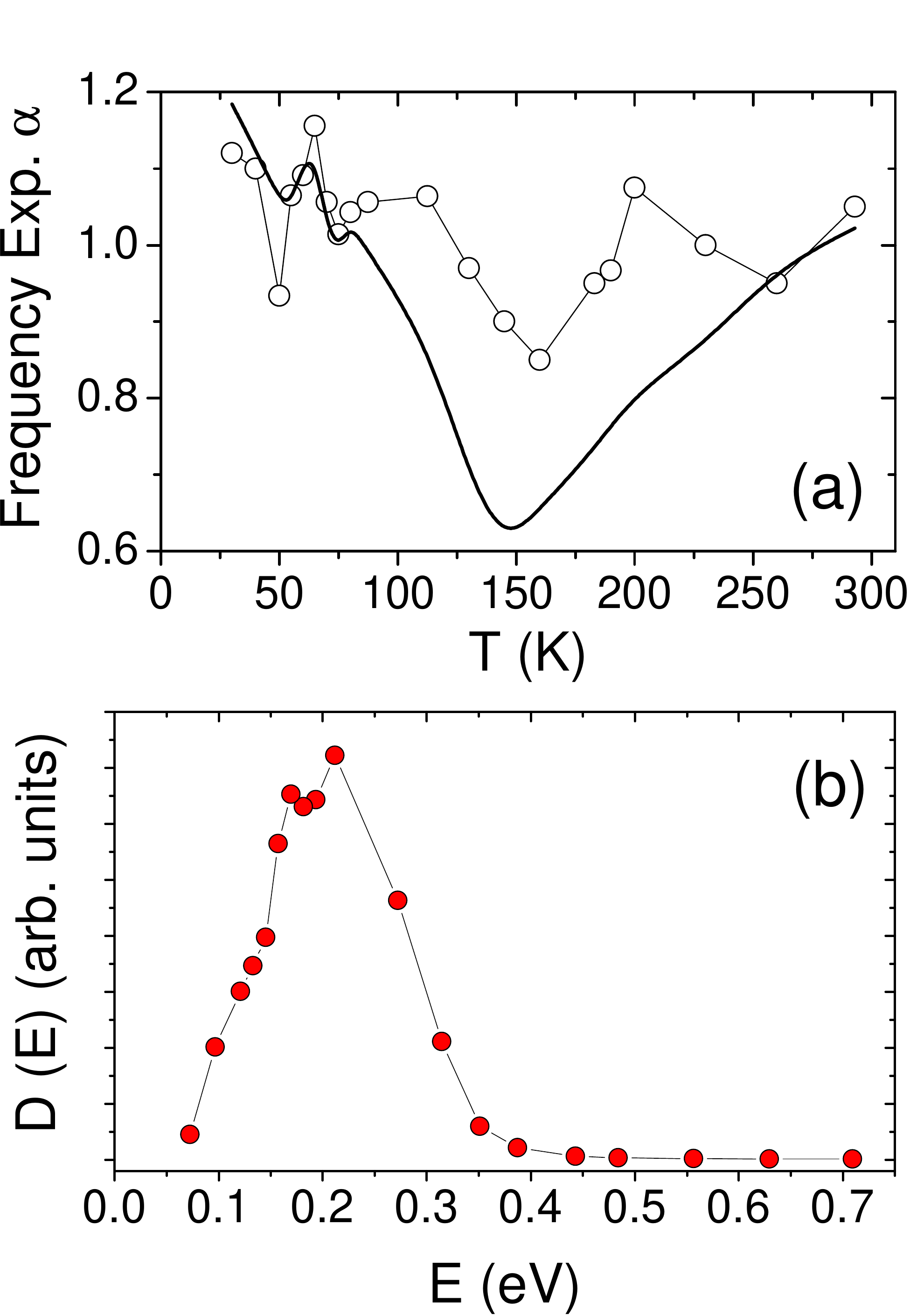}
\caption[]{(Color online). (a) Frequency exponent $\alpha(T)$ of the $1/f^\alpha$ noise spectra. Line is the prediction from the DDH model Eq.\,\ref{DDHeq3} without considering a temperature dependent function $g(T)$. (b) Energy distribution calculated from the data in Fig.\,\ref{figure6} using Eq.\,\ref{DDHeq2}, again with $g(T) = const$.}\label{figure7}
\end{figure}
discussed above, for ambient-pressure \kCl\ we observe the expected semiconducting (insulating) behavior for both the out-of-plane (not shown) and in-plane resistances. Following the analysis in \cite{Kagawa2004}, we get an estimate for the effective charge gap from a fit to the data at low temperatures $R \propto \Delta/T$ yielding $\Delta \sim 300$\,K. While the resistance shows a smooth behavior with no apparent features, the electronic fluctuations exhibit a pronounced shoulder in the resistance noise power spectral density $S_R(T)$ corresponding to a broad maximum in the normalized noise $S_R/R^2(T)$ at around 90\,K, see Fig.\,\ref{figure6}. In Fig.\,\ref{figure7} (a) we show the DDH analysis as discussed for \kClstar, without considering a function $g(T)$ taking into account a possible temperature dependence of the number and/or strength of the fluctuators. Again, we find that the frequency exponents are distributed closely around $\alpha = 1$ and that the predicted curve reproduces the general course of the measured $\alpha(T)$ values, which allows us to  extract the energy distribution $D(E)$ from the noise data in Fig.\,\ref{figure6}. The latter is depicted in Fig.\,\ref{figure7} (b). The analysis yields a maximum at an energy of $\sim 212$\,meV. As above, the substantially enhanced noise level around 90\,K most likely corresponds to the thermal excitations of the ethylene endgroups of the ET molecules causing enhanced low-frequency fluctuations of the charge carriers. Further studies on different crystals of the title compound are necessary to identify possible sample-to-sample variations and investigate the significance of the finestructure in $D(E)$ observed at 170\,meV.

\section{Conclusion}
Resistance noise spectroscopy studies on bulk single crystals of the quasi-2D organic Mott system \Cl\ revealed $1/f^\alpha$-type spectra in the temperature range up to room temperature. The observed noise level is extremely enhanced compared to homogeneous semiconductors and metals and may be due to an inhomogeneous current distribution inherent to the present molecular conductors. The temperature dependence of the noise is remarkably well described by the Dutta-Dimon-Horn random fluctuation model. We find that the model's assumptions of independent and thermally-activated fluctuators apply for to systems studied here. Our results for the system under a moderate pressure suggest that the overall noise is temperature dependent, i.e.\ the number and/or the strength of the switching entities increase with decreasing temperature according to a $T^{-3/2}$ power law. The same scattering mechanism seems to describe the resistivity of the sample. The distribution of activation energies $D(E)$ of the fluctuators causing the nearly $1/f$ noise can be extracted from the model. The most striking feature is a pronounced maximum in $D(E)$ at around 230\,meV, which corresponds to the activation energy of the conformational motion of the ET molecules' terminal ethylene groups. It is interesting to note that the peak in the noise has no correspondence in the actual resistivity measurements. The observation of the ethylene endgroups' glassy energy signature, which is well known from thermodynamic studies, in the electronic noise properties demonstrates that fluctuation spectroscopy provides a promising novel access to the microscopic transitions and dynamical molecular properties of the present materials. 

\begin{acknowledgments}
The work is supported by the Deutsche Forschungsgemeinschaft (DFG) through the Emmy Noether program. J.M. gratefully acknowledges fruitful discussions with Dr.\ S. Wirth. Work at Argonne National Laboratory is sponsored by the U.S. Department of Energy, Office of Basic Energy Sciences, Division of Materials Sciences, under Contract No.\ DE-AC02-06CH11357. 
\end{acknowledgments}


\end{document}